\documentclass{article}
\usepackage{authblk}
\usepackage{graphicx} 
\usepackage{amsmath}
\usepackage{amssymb}
\usepackage{braket}
\usepackage{url}
\usepackage[hidelinks]{hyperref}
\usepackage{listings}
\usepackage{xcolor}

\definecolor{codegreen}{rgb}{0,0.6,0}
\definecolor{codegray}{rgb}{0.5,0.5,0.5}
\definecolor{codepurple}{rgb}{0.58,0,0.82}
\definecolor{backcolour}{rgb}{0.95,0.95,0.92}

\lstdefinestyle{mystyle}{
    backgroundcolor=\color{backcolour},   
    commentstyle=\color{codegreen},
    keywordstyle=\color{magenta},
    numberstyle=\tiny\color{codegray},
    stringstyle=\color{codepurple},
    basicstyle=\ttfamily\footnotesize,
    breakatwhitespace=false,         
    breaklines=true,                 
    captionpos=b,                    
    keepspaces=true,                 
    numbers=left,                    
    numbersep=5pt,                  
    showspaces=false,                
    showstringspaces=false,
    showtabs=false,                  
    tabsize=2
}

\usepackage{biblatex} 
\usepackage{layout}

\title{Pseudospectral discretization of the time-dependent Schrödinger equation}
\author{Håkon Emil Kristiansen}
\author{Einar Aurbakken}
\affil{Hylleraas Centre for Quantum Molecular Sciences, Department of Chemistry, University of Oslo, P.O. Box 1033 Blindern, N-0315 Oslo, Norway
}%

\begin{document}

\date{\today}
\maketitle

\tableofcontents

\newpage

\section{Introduction}
The purpose of this document is to describe the solution and implementation of the time-independent and time-dependent Schrödinger using pseudospectral methods. 
Currently, the description is for single particle systems interacting with a classical electromagnetic field in spherical coordinates.

\section{The Schrödinger equation}
We want to solve the time-dependent Schrödinger (TDSE) equation, expressed in atomic units $\hbar = |e| = m_e = 4 \pi \epsilon_0 = 1$,
\begin{equation}
    i \dot{\Psi}(\mathbf{r},t) = \hat{H}(\mathbf{r},t)\Psi(\mathbf{r},t), \label{TDSE} \\
\end{equation}
given some initial wavefunction/state 
\begin{equation}
    \Psi(\mathbf{r}, t_0) = \Psi_0.
\end{equation}

The Hamiltonian (in atomic units) of an electron confined by a potential $V(\mathbf{r})$ interacting with a time-dependent (classical) electromagnetic field given by the vector potential $\mathbf{A}(\mathbf{r},t)$ in the Coulomb gauge ($\nabla \cdot \mathbf{A}(\mathbf{r},t) = 0$), is given by~\cite{joachain2012atoms}

\begin{align}
    \hat{H}(\mathbf{r}, t) &= \frac{1}{2} \left( \hat{p}+\mathbf{A}(\mathbf{r},t) \right)^2 + V(\mathbf{r}) \\
    &= -\frac{1}{2}\nabla^2 + V(\mathbf{r}) + \mathbf{A}(\mathbf{r},t)\cdot \hat{p} + \frac{1}{2}\mathbf{A}(\mathbf{r},t)^2 \\
    &= \hat{H}_0(\mathbf{r}) + \hat{H}_{\text{int}}(\mathbf{r},t),
\end{align}
where we have defined the field-free and interaction Hamiltonian,
\begin{align}
    \hat{H}_0(\mathbf{r}) &\equiv -\frac{1}{2}\nabla^2 + V(\mathbf{r}), \\
    \hat{H}_{\text{int}}(\mathbf{r},t) &\equiv \mathbf{A}(\mathbf{r},t)\cdot \hat{p} + \frac{1}{2}\mathbf{A}(\mathbf{r},t)^2.
\end{align}
Typically, the initial state is taken as the ground state of the field-free Hamiltonian, which can be determined by solving the time-independent Schrödinger equation
\begin{equation}
    \hat{H}_0(\mathbf{r}) \Psi_k(\mathbf{r}) = \epsilon_k \Psi_k(\mathbf{r}), \ \ k=1,2,3,\cdots \label{TISE}.
\end{equation}

It is common to ignore the spatial dependence of the field, i.e., $\mathbf{A}(\mathbf{r},t) \approx \mathbf{A}(t)$, and is referred to as the \textit{electric dipole approximation}, and the interaction operator reduces to
\begin{equation}
    \hat{H}_{\text{int}}(\mathbf{r},t) \equiv \mathbf{A}(t)\cdot \hat{p} + \frac{1}{2}\mathbf{A}(t)^2.
\end{equation}
One can eliminate the $\mathbf{A}^2$ term by extracting a time-dependent phase from the wavefunction and the resulting interaction operator~\cite{joachain2012atoms}
\begin{equation}
    \hat{H}^V_{\text{int}}(\mathbf{r},t) \equiv \mathbf{A}(t)\cdot \hat{p},
\end{equation}
is said to be in the \textit{velocity gauge}. The interaction operator can be transformed/simplified by a gauge transformation. In particular,  the \textit{length gauge} form of the interaction operator is given by
\begin{equation}
    \hat{H}^L_{\text{int}}(\mathbf{r},t) \equiv \mathcal{E}(t)\cdot \mathbf{r},
\end{equation}
where 
\begin{equation}
    \mathcal{E}(t) = -\frac{\partial }{\partial t} \mathbf{A}(t).
\end{equation}

\subsection{Spherical coordinates}
In spherical coordinates (see appendix \ref{sec:sph_coords}), we parametrize the wavefunction as
\begin{equation}
    \Psi(\mathbf{r},t) = \sum_{l=0}^{\infty} \sum_{m=-l}^l r^{-1}u_{l,m}(r,t) Y_{l,m}(\Omega) \label{sph_wf},
\end{equation}
where $Y_{l,m}(\Omega)$ are the spherical harmonics using the convention described in appendix \ref{sec:sph_harmonics}.
The action of the Laplacian on the wavefunction is given by
\begin{equation}
    \nabla^2 \Psi(\mathbf{r}) = \sum_{l,m} r^{-1} \left( \frac{d^2 }{dr^2} - \frac{l(l+1)}{r^2}\right) u_{l,m}(r)  Y_{l,m}(\Omega),
\end{equation}
where we have defined 
\begin{equation}
    \Delta_l \equiv \Delta_l\left(r, \frac{d^2}{d r^2} \right) = \frac{d^2 }{dr^2} - \frac{l(l+1)}{r^2}.
\end{equation}

\subsubsection{TDSE}
Inserting the wavefunction Ansatz (\ref{sph_wf}) into the TDSE (\ref{TDSE}), multiplying through with $r$, $Y_{l,m}^*(\Omega)$ and integrating over $\Omega$ yields 
\begin{equation}
    i \dot{u}_{l,m}(r,t) = -\frac{1}{2}\Delta_lu_{l,m}(r,t) + r\int Y_{l,m}^*(\Omega)\left(V(r,\Omega)+\hat{H}_{\text{int}}(r,\Omega, t) \right)\Psi(r,\Omega,t) d\Omega, \label{radial_TDSE}
\end{equation}
where $d\Omega = \sin \theta d\theta d\phi$. 

\subsubsection{TISE}
If we expand each eigenfunction $\Psi_k(\mathbf{r})$ as
\begin{equation}
    \Psi_k(r,\Omega) = \sum_{l=0}^\infty \sum_{m=-l}^l r^{-1} u^{(k)}_{l,m}(r) Y_{l,m}(\Omega),
\end{equation}
insertion into the TISE~(\ref{TISE}), multiplying through with $r$, $Y_{l,m}^*(\Omega)$ and integrating over $\Omega$ yields a set of eigenvalue equations for the radial functions $u^{(k)}_{l,m}(r)$ given by
\begin{equation}
    -\frac{1}{2}\Delta_lu^{(k)}_{l,m}(r) + r\int Y_{l,m}^*(\Omega)V(r,\Omega)\Psi(r,\Omega) d\Omega = \epsilon_k u^{(k)}_{l,m}(r). \label{radial_TISE}
\end{equation}
Notice that for angular-dependent potentials, the second term on the left hand side generally leads to couplings of/in angular momenta/$(l,m)$, and the eigenfunctions $\Psi_k(r, \Omega)$ are not states of definite angular momentum (not eigenfunctions $\hat{L}^2$ and $\hat{L}_z$).  

If the potential is spherically symmetric, i.e.,  $V(r,\Omega) = V(r)$, the eigenfunctions are states with definite angular momentum and we can assume that 
\begin{equation}
    \Psi_k(\mathbf{r}) = \Psi_{(n,l,m)}(r,\Omega) = r^{-1} \psi_{n,l}(r)Y_{l,m}(\Omega),
\end{equation}
where $k=(n,l,m)$ and the TISE for the radial functions $\psi_{n,l}(r)$ reduce to
\begin{equation}
    -\frac{1}{2}\Delta_l\psi_{n,l}(r) + V(r)\psi_{n,l}(r) = \epsilon_{n,l} \psi_{n,l}(r). \label{radial_sph_symmetric_TISE}
\end{equation}

In order to solve equations (\ref{radial_TDSE}), (\ref{radial_TISE}) and/or (\ref{radial_sph_symmetric_TISE}) numerically, we have to choose a cutoff $l_{\text{max}}$ for the angular momentum and discretize the radial coordinate. For this, we will use the pseudospectral method, also referred to as the discrete variable representation (DVR), which we describe in Section~\ref{pseudospectral}.

\section{Integration of equations of motion}
\subsection{The time evolution operator}
The solution of the TDSE~(\ref{TDSE}) can formally be expressed as~\cite{joachain2012atoms, ullrich2011time}
\begin{equation}
    \psi(t+\Delta t) = \hat{U}(t+\Delta t, t) \psi(t), \label{TDSE_evolution_operator}
\end{equation}
where $\hat{U}(t+\Delta t, t)$ is the time evolution operator taking the wavefunction from time $t$ to $t+\Delta t$, with $\Delta t$ being the time step.

For a sufficiently small time step, the time evolution operator can be approximated as~\cite{joachain2012atoms}
\begin{equation}
    \hat{U}(t_n+\Delta t, t_n) = \exp (-i\Delta t \hat{H}(t_n+\Delta t/2)) \label{evolution_operator_midpoint},
\end{equation}
where the (time-dependent) Hamiltonian is evaluated at the midpoint of the time interval $t_n + \Delta t/2$. Usually, it is computationally too expensive to evaluate the exact exponential in Eq.~(\ref{evolution_operator_midpoint}), and one has to introduce further approximations.

\subsection{The Crank-Nicolson method}
\label{sec:crank_nicolson}
In the Crank-Nicolson method, the exponential in Eq.~(\ref{evolution_operator_midpoint}) is approximated as
\begin{equation}
    \exp (-i\Delta t \hat{H}^{n+\frac{1}{2}}) \approx \left(1+\frac{i\Delta t}{2} \hat{H}^{n+\frac{1}{2}}\right)^{-1}\left(1-\frac{i\Delta t}{2} \hat{H}^{n+\frac{1}{2}}\right),
\end{equation}
where $1$ is the identity operator and $\hat{H}^{n+\frac{1}{2}} \equiv \hat{H}(t_n+\Delta/2)$. Then, Eq.~(\ref{TDSE_evolution_operator}) can be written as
\begin{equation}
    \left(1+\frac{i \Delta t}{2} \hat{H}^{n+\frac{1}{2}} \right) \psi^{n+1} = \left(1-\frac{i \Delta t}{2} \hat{H}^{n+\frac{1}{2}} \right) \psi^{n} \label{CN_operator_form},
\end{equation}
where $\psi^n = \psi(t_n+\Delta t)$. 

In the forthcoming discussion, we assume that the wavefunction is expressed in a finite basis and that the operators $1$ and $\hat{H}(t)$ are represented by finite-dimensional matrices $I$ and $H(t)$ in this basis.  Then, by Eq.~(\ref{CN_operator_form}), the wavefunction at the next time step the solution of the linear equation
\begin{equation}
     A^{n+\frac{1}{2}} \psi^{n+1} = b^n, 
\end{equation}
where
\begin{align}
    A^{n+\frac{1}{2}} &=  I+\frac{i \Delta t}{2} H^{n+\frac{1}{2}}, \\
    b^n &= \left( I-\frac{i \Delta t}{2} H^{n+\frac{1}{2}} \right) \psi^{n}
\end{align}
In practice, it is often too expensive to solve this equation directly either by explicitly inverting $A^{n+\frac{1}{2}}$  or by Gauss-elimination, and one has to use an iterative method. Note also that $A^{n+\frac{1}{2}}$ is non-Hermitian
\begin{equation}
    (A^{n+\frac{1}{2}})^\dagger = I-\frac{i \Delta t}{2} H^{n+\frac{1}{2}} \neq A^{n+\frac{1}{2}},
\end{equation}
so we have to use a method that can handle non-Hermitian problems.

The BiConjugate Gradient Stabilized method (Bi-CGSTAB)~\cite{barrett1994templates} is one such option and is available/implemented in the \texttt{SciPy} sparse library~\footnote{\href{https://docs.scipy.org/doc/scipy/reference/generated/scipy.sparse.linalg.bicgstab.html}{scipy.sparse.linalg.bicgstab.html}}. 
The "bicgstab" routine solves the linear system
\begin{equation}
    Ax = b, \label{lin_system}
\end{equation}
provided a function for computing the matrix-vector product $y = Ax$, the right hand side $b$, an initial guess $x_0$ for the solution, and terminates if/when  
\begin{equation}
\lVert b - Ax_k \rVert \leq \max (r_{\text{tol}} \lVert b \rVert, a_{\text{tol}}),
\end{equation}
where $k$-th iterate $x_k$ is the approximate solution to/of the linear system~(\ref{lin_system}).
The parameters $a_{\text{tol}}$ and $r_{\text{tol}}$ are user specified and the default values are $a_\text{tol} = 0.0$ and $r_{\text{tol}} = 10^{-5}$. 

Additionally, a preconditioner $M$ can be provided to (possibly) accelerate convergence. A preconditioner is a matrix that should be an approximation to the inverse of $A$, i.e., $M \approx A^{-1}$. Multiplying with $M$ on both sides of Eq.~(\ref{lin_system}) we obtain a new linear system
\begin{equation}
    M y = \Tilde{b},
\end{equation}
where $\Tilde{b} = Mb$. If $M$ is chosen properly, this new system should be easier/better conditioned than the original problem and (hopefully) require fewer iterations to solve. 
The implementation of the matrix-vector product $Ax$ and preconditioner $M$ are specific to the parametrization/coordinate system we use. 

\section{The pseudospectral method}\label{pseudospectral}
\subsection{General description}
In the pseudospectral method, a function $f(x)$ is approximated as a linear combination
\begin{equation}
    f(x) \simeq f_N(x) = \sum_{j=0}^N f(x_j) g_j(x)
\end{equation}
of cardinal functions $g_j(x)$ where the expansion coefficients are values of $f(x)$ at a set of grid points $\{x_j\}_{j=0}^N$. 

For Lobatto-type grids, the interior grid points $\{ x_j \}_{j=1}^{N-1}$ are defined as the zeros of $P_N^\prime(x)$, where $P_N(x)$ is a $N$-th order polynomial,
\begin{equation}
    P_N^\prime(x_j) = 0, \ \ j=1,\cdots,N-1,
\end{equation}
while the end (boundary) points $x_0$ and $x_N$ are given. 
The type of polynomial to choose depends on the problem, with the most common choices being classical orthogonal polynomials such as Chebyshev, Legendre, Laguerre, and Hermite polynomials. It is also possible to construct pseudospectral grids from other functions, with one example being the Coulomb wave function DVR method described in Ref.~\cite{peng2006_coulombwave_dvr}.

The cardinal functions $g_j(x)$ have the property that
\begin{equation}
    g_j(x_i) = \delta_{ij}, \label{delta_property_g}
\end{equation}
where $\delta_{ij}$ is the Kronecker delta, and can be written as 
\begin{equation}
    g_j(x) = -\frac{1}{N(N+1)P_N(x_j)}\frac{(1-x^2)P_N^\prime(x)}{x-x_j}.
\end{equation}

The integral of a function is approximated by quadrature
\begin{equation}
    \int_{x_0}^{x_N} f(x) dx \simeq \sum_{j=0}^N w_j f(x_j), \label{quadrature_rule}
\end{equation}
where $w_j$ are the quadrature weights that depend on the underlying choice of polynomial $P_N(x)$. 

\subsection{Gauss-Legendre-Lobatto}
In the Gauss-Legendre-Lobatto (GLL) pseudospectral method~\cite{boyd2001chebyshev}, $P_N(x)$ is taken as the $N$-th order Legendre polynomial. The end points are $x_0 = -1$ and $x_N=1$, while the interior grid points must be found as the roots 
\begin{equation}
    P_N^\prime(x_i) = 0,
\end{equation}
and the quadrature weights are given by
\begin{equation}
    w_i = \frac{2}{N(N+1)P_N(x_i)^2}.
\end{equation}
The grid points and weights can be determined using the \texttt{polynomial}~\footnote{\url{https://numpy.org/doc/stable/reference/routines.polynomials.html}} package in \texttt{NumPy} as demonstrated in Listing \ref{GLL_nodes_and_weights}.

Furthermore~\cite{boyd2001chebyshev, cinal2020highly}, 
\begin{align}
    \frac{d g_j}{dx} \Bigr|_{x=x_i} &= g_j^\prime(x_i) = \tilde{g}_j^\prime(x_i) \frac{P_N(x_i)}{P_N(x_j)}, \\
    \tilde{g}^\prime_j(x_i) &= 
    \begin{cases}
    \frac{1}{4}N(N+1), \ \ i=j=0, \\
    -\frac{1}{4}N(N+1), \ \ i=j=N, \\
    0, \ \ i=j \text{ and } 1 \leq j \leq N-1, \\
    \frac{1}{x_i-x_j}, \ \ i \neq j,
    \end{cases}
    \label{dg_dx_dg_tilde}
\end{align}
and for $i,j = 1,\cdots,N-1$
\begin{align}
    \frac{d^2 g_j}{dx^2} \Bigr|_{x=x_i} &= g_j^{\prime \prime}(x_i) = \tilde{g}_j^{\prime \prime}(x_i) \frac{P_N(x_i)}{P_N(x_j)}, \label{ddg_ddx}  \\
    \tilde{g}_j^{\prime \prime}(x_i) &= 
    \begin{cases}
    -\frac{1}{3} \frac{N(N+1)}{(1-x_i^2)}, i = j,\\
    -\frac{2}{(x_i-x_j)^2}, i \neq j.
    \end{cases} \label{ddg_ddx_ddg_tilde}
\end{align}




\begin{lstlisting}[language=Python, caption=Code example using \texttt{NumPy} to compute Gauss-Legendre-Lobatto grid points (nodes) and quadrature weights., label=GLL_nodes_and_weights]
import numpy as np
N = 30 
PN = np.zeros(N+1)
PN[-1] = 1
dPN_dx = np.polynomial.legendre.legder(PN)

x = np.zeros(N+1)
x[0] = -1
x[N] = 1
x[1:-1] = np.polynomial.legendre.legroots(dPN_dx)

PN_x = np.polynomial.legendre.legval(x, PN)
weights = 2 / (N * (N + 1)*PN_x**2)
\end{lstlisting}

\section{The pseduospectral method applied to the Scrödinger equation in spherical coordinates}
\subsection{The time-independent Schrödinger equation}
First, we consider the time-independent Schrödinger equation for a spherically symmetric potential given by Eq.~(\ref{radial_sph_symmetric_TISE}).

The GLL grid points are defined on the interval $[-1,1]$, and 
since we use spherical coordinates, we have to map the grid points $x_i \in [-1,1]$ into radial points $r(x): [-1,1] \rightarrow [0, r_{\text{max}}]$, where $r_{\text{max}}$ is a chosen (finite) cut-off for the radial grid. 
In that case, $\psi_{n,l}(r) = \psi_{n,l}(r(x))$.

The (rational) mapping
\begin{equation}
    r(x) = L \frac{1+x}{1-x+\alpha}, \ \ \alpha = \frac{2L}{r_{\text{max}}} \label{x_to_r},
\end{equation}
has often been used in earlier work~\cite{Wang_MultiphotonDetachment_1994, Greenman_TDCIS_2010, cinal2020highly}. 
The parameters $L$ and $\alpha$ control the length of the grid and the density of points near the boundaries $r=0$ and $r = r_{\text{max}}$. 

Another option is to use the linear mapping given by~\cite{rescigno2000numerical} 
\begin{equation}
    r(x) = \frac{r_{\text{max}}}{2}(x+1). \label{linear_x_to_r}
\end{equation}

To use the Gauss-Legendre-Lobatto pseudospectral method, we must first formulate Eq.~(\ref{radial_sph_symmetric_TISE}) with respect to $x$. Using the chain rule (see \ref{subsec:chain_rule}) the radial TISE becomes
\begin{equation}
    \left(-\frac{1}{2} \left( \frac{1}{\dot{r}(x)^2} \frac{d^2 }{dx^2} - \frac{\ddot{r}(x)}{\dot{r}(x)^3} \frac{d }{dx}  \right) + \frac{l(l+1)}{2r(x)^2} + V(r(x) \right) \psi_{n,l}(r(x)) = \epsilon_{n,l} \psi_{n,l}(r(x)). \label{SE_of_x},
\end{equation}
where 
\begin{equation}
    \dot{r}(x) \equiv \frac{d}{dx}r(x).
\end{equation}
Discretization of Eq.~(\ref{SE_of_x}) will result in an unsymmetric (non-Hermitian) eigenvalue problem due to the presence of the derivative operator $d/dx$ which is anti-symmetric.

To end up with a symmetric (Hermitian) eigenvalue problem, we introduce the scaling~\cite{Wang_MultiphotonDetachment_1994} 
\begin{equation}
    \psi_{n,l}(r(x)) = \dot{r}(x)^{-1/2}f_{n,l}(x). \label{symm_scaling}
\end{equation}
Insertion into Eq.(\ref{SE_of_x}) yields (see \ref{subsec:symmetrization}) 

\begin{equation}
\left(-\frac{1}{2} \frac{1}{\dot{r}(x)} \frac{d^2}{dx^2} \frac{1}{\dot{r}(x)}  + \tilde{V}_l(r(x)) \right) f_{n,l}(x) = \epsilon_{n,l} f_{n,l}(x) \label{SE_of_x_f}, 
\end{equation}
where the operator $\frac{1}{\dot{r}(x)} \frac{d^2}{dx^2} \frac{1}{\dot{r}(x)}$ is symmetric and  we have defined
\begin{equation}
    \tilde{V}_l(r(x)) \equiv V(r(x)) + \frac{l(l+1)}{2r(x)^2} + \frac{2\dddot{r}(x)\dot{r}(x)-3\ddot{r}(x)^2}{4\dot{r}(x)^4} \label{func:tilde_Vl}.
\end{equation}
Note that for the mappings given by Eqs.~(\ref{x_to_r}, \ref{linear_x_to_r}), $\tilde{V}_l(r(x)) = V(r(x)) + \frac{l(l+1)}{2r(x)^2}$, since
\begin{equation}
    2\dddot{r}(x)\dot{r}(x)-3\ddot{r}(x)^2 = 0.
\end{equation}

In order to discretize Eq.~(\ref{SE_of_x_f}) we now expand $f_{n,l}(x)$ and $f_{n,l}(x)/\dot{r}(x)$ as 
\begin{align}
    f_{n,l}(x) &= \sum_{j=0}^N f_{n,l}(x_j) g_j(x), \\
    \frac{f_{n,l}(x)}{\dot{r}(x)} &= \sum_{j=0}^N \frac{f_{n,l}(x_j)}{\dot{r}(x_j)} g_j(x).
\end{align}
Inserting these expansions into Eq.~(\ref{SE_of_x_f}) we have that 
\begin{equation}
     \sum_{j=1}^{N-1} \left(-\frac{1}{2} \frac{1}{\dot{r}(x)} \frac{f_{n,l}(x_j)}{\dot{r}(x_j)} g^{\prime \prime}_j(x) + \tilde{V}_l(r(x)) f_{n,l}(x_j) g_j(x) \right) = \epsilon_{n,l} \sum_{j=1}^{N-1} f_{n,l}(x_j) g_j(x)
\end{equation}
Next, we multiply through with $g_i(x)$ and integrate over $x$ 
\begin{align}
     &\sum_{j=1}^{N-1} \left(-\frac{1}{2}  \frac{f_{n,l}(x_j)}{\dot{r}(x_j)} \int \frac{g_i(x)}{\dot{r}(x)} g^{\prime \prime}_j(x) dx +  f_{n,l}(x_j) \int g_i(x) \tilde{V}_l(r(x)) g_j(x) dx \right) \nonumber \\ 
     = \epsilon_{n,l} &\sum_{j=1}^{N-1} f_{n,l}(x_j) \int g_i(x) g_j(x) dx.
\end{align}

The integrals are approximated with the quadrature rule (\ref{quadrature_rule}). Also, recall that $g_j^{\prime \prime}(x_i)$ given by Eq.~(\ref{ddg_ddx}) is defined only for the interior grid points. Thus, we find that
\begin{align}
    \int g_i(x) g_j(x) dx &\approx \sum_{m=0}^N g_i(x_m) g_j(x_m) w_m = \sum_{m=0} w_m \delta_{i, m} \delta_{j,m} = w_i \delta_{i,j}, \\
    \int g_i(x) \tilde{V}_l(r(x)) g_j(x) dx &\approx \sum_{m=0}^N g_i(x_m) \tilde{V}_l(r(x_m)) g_j(x_m) w_m = w_i \tilde{V}_l(r(x_i)) \delta_{i,j}, \\
    \int \frac{g_i(x)}{\dot{r}(x)} g^{\prime \prime}_j(x) dx & \approx \sum_{m=1}^{N-1} \frac{g_i(x_m)}{\dot{r}(x_m)} g^{\prime \prime}_j(x_m) = w_i \frac{g^{\prime \prime}_j(x_i)}{\dot{r}(x_i)},  \ \ i=1,\cdots,N-1,
\end{align}
where we have used the property (\ref{delta_property_g}).
Then, we have that (for the interior grid points),
\begin{equation}
     \sum_{j=1}^{N-1} \left(-\frac{1}{2}  \frac{f_{n,l}(x_j)}{\dot{r}(x_j)} w_i \frac{g^{\prime \prime}_j(x_i)}{\dot{r}(x_i)} +  f_{n,l}(x_j) w_i \tilde{V}_l(r(x_i)) \delta_{i,j} \right) = \epsilon_{n,l} \sum_{j=1}^{N-1} f_{n,l}(x_j) w_i \delta_{i,j}. 
\end{equation}
Using Eq.~(\ref{ddg_ddx_ddg_tilde}) we can write this equation as (notice that the weights $w_i$ cancels)
\begin{equation}
     \sum_{j=1}^{N-1} \left(-\frac{1}{2}  \frac{\tilde{g}^{\prime \prime}_j(x_i) P_N(x_i)}{\dot{r}(x_i) \dot{r}(x_j)} \frac{f_{n,l}(x_j)} {P_N(x_j)} \right) +  f_{n,l}(x_i) \tilde{V}_l(r(x_i))  = \epsilon_{n,l}  f_{n,l}(x_i). 
\end{equation}
Furthermore, dividing through with $P_N(x_i)$, we have that 
\begin{equation}
     \sum_{j=1}^{N-1} \left(-\frac{1}{2}  \frac{\tilde{g}^{\prime \prime}_j(x_i)}{\dot{r}(x_i) \dot{r}(x_j)} \tilde{f}_{n,l}(x_j) \right) +   \tilde{V}_l(r(x_i))\tilde{f}_{n,l}(x_i)  = \epsilon_{n,l}  \tilde{f}_{n,l}(x_i) \label{discrete_tise_of_x},
\end{equation}
where we have defined 
\begin{align}
    \tilde{f}_{n,l}(x_i) \equiv \frac{f_{n,l}(x_i)}{P_N(x_i)}.
\end{align}
Eq.~(\ref{discrete_tise_of_x}) can be written (on matrix form) as
\begin{equation}
    H_l \mathbf{\tilde{f}}_{n,l} = \epsilon_{n,l}\mathbf{\tilde{f}}_{n,l},
\end{equation}
where we have defined
\begin{align}
    H_l &\equiv T_l + V, \\
    T_l &\equiv -\frac{1}{2}D_2 + \left[\frac{l(l+1)}{2(r(x_i)} \delta_{i,j} \right], \label{Tl_matrix} \\
    D_{2} &\equiv \left[ \frac{\tilde{g}^{\prime \prime}_j(x_i)}{\dot{r}(x_i) \dot{r}(x_j)} \right], \\ 
    V &\equiv \left[ \left(V(x_i)+\frac{2\dddot{r}(x_i)\dot{r}(x_i)-3\ddot{r}(x_i)^2}{4\dot{r}(x_i)^4}\right) \delta_{i,j}\right],
\end{align}
and 
\begin{equation}
    \mathbf{\tilde{f}}_{n,l} \equiv 
    \begin{bmatrix}
    f_{n,l}(x_1) \\
    f_{n,l}(x_2) \\
    \vdots \\
    f_{n,l}(x_{N-1})
    \end{bmatrix}
\end{equation}
Here we have used the notation
\begin{equation}
    A = [a_{i,j}],
\end{equation}
to denote a matrix $A$ with matrix elements $A_{i,j} = a_{i,j}$. 
The matrix $H_l$ is symmetric/Hermitian and can be diagonalized by an eigensolver tailored for symmetric/Hermitian matrices such as the \texttt{eigh}~\footnote{\url{https://numpy.org/doc/stable/reference/generated/numpy.linalg.eigh.html}} function in \texttt{NumPy}. 

\subsection{The time-dependent Schrödinger equation}
\label{sph_discrete_tise}
Consider the radial TDSE given by Eq.~(\ref{radial_TDSE}). 

Assume that the potential $V$ is spherically symmetric and that the interaction is in the dipole approximation in the velocity gauge 
\begin{equation}
    \hat{H}^V_{\text{int}}(r,\Omega,t) = A(t) \mathbf{u} \cdot \mathbf{p},
\end{equation}
where $\mathbf{u}$ is a polarization vector of unit length and 
\begin{align}
    \mathbf{p} &= p_x \mathbf{e}_x + p_y \mathbf{e}_y + p_z \mathbf{e}_z \\
    &= -i\left( \frac{\partial}{\partial x} \mathbf{e}_x + \frac{\partial}{\partial y} \mathbf{e}_y +\frac{\partial}{\partial z} \mathbf{e}_z \right),
\end{align}
where the Cartesian derivative operators in spherical coordinates are given by Eqs.~(\ref{sph_dx})-(\ref{sph_dz}). Furthermore, the Cartesian derivative operators can be written on the form  
\begin{equation}
    \frac{\partial}{\partial K} = \alpha_K(\Omega) \frac{\partial}{\partial r} + \frac{1}{r} \beta_K(\Omega), \ \ K\in \{x,y,z \}, 
\end{equation}
where 
\begin{align}
    \alpha_K &= 
    \begin{cases}
    \cos \phi \sin \theta, \ \ K=x, \\
    \sin \phi \sin \theta, \ \ K=y, \\
    \cos \theta, \ \ K=z,
    \end{cases} \\
    \beta_K &= 
    \begin{cases}
    -\frac{\sin \phi}{\sin \theta}\frac{\partial}{\partial \phi} + \cos \phi \cos \theta \frac{\partial}{\partial \theta}, \ \ K=x, \\
    \frac{\cos \phi}{\sin \theta}\frac{\partial}{\partial \phi} + \sin \phi \cos \theta \frac{\partial}{\partial \theta}, \ \ K=y, \\
    -\sin \theta \frac{\partial}{\partial \theta}, \ \ K=z.
    \end{cases}
\end{align}
In addition to the above assumptions, if the field is linearly polarized, $\mathbf{u} = \mathbf{e}_K$, the radial TDSE reduces to
\begin{align}
    i \dot{u}_{l,m}(r,t) = &-\frac{1}{2}\nabla_l u_{l,m}(r,t) + V(r)u_{l,m}(r,t) \nonumber \\ 
    &-iA(t)r\int Y_{l,m}^*(\Omega) \left( \alpha_K(\Omega) \frac{\partial}{\partial r} + \frac{1}{r} \beta_K(\Omega) \right) \Psi(r,\Omega,t) d\Omega.
\end{align}

Using that 
\begin{equation}
    \Psi(r,\Omega,t) = \sum_{l^\prime,m^\prime} r^{-1} u_{l^\prime, m^\prime}(r,t) Y_{l^\prime, m^\prime}(\Omega),
\end{equation}
the interaction/coupling term can be written as
\begin{align}
    &r\int Y_{l,m}^*(\Omega) \left( \alpha_K(\Omega) \frac{\partial}{\partial r} + \frac{1}{r} \beta_K(\Omega) \right) \Psi(r,\Omega,t) d\Omega \nonumber \\
    &=r\sum_{l^\prime,m^\prime} \left(  \alpha^K_{(l,m),(l^\prime, m^\prime)} \frac{\partial }{\partial r}\left(\frac{u_{l^\prime, m^\prime}(r,t)}{r} \right)   + \beta^K_{(l,m),(l^\prime, m^\prime)} \frac{u_{l^\prime, m^\prime}(r,t)}{r^2} \right) \nonumber \\
    &=r\sum_{l^\prime,m^\prime} \left(  \alpha^K_{(l,m),(l^\prime, m^\prime)} \left(\frac{1}{r}\frac{\partial u_{l^\prime, m^\prime}(r,t)}{\partial r} - \frac{u_{l^\prime, m^\prime}(r,t)}{r^2} \right)   + \beta^K_{(l,m),(l^\prime, m^\prime)} \frac{u_{l^\prime, m^\prime}(r,t)}{r^2} \right) \nonumber \\
    &= \sum_{l^\prime,m^\prime} \left( \alpha^K_{(l,m),(l^\prime, m^\prime)} \frac{\partial u_{l^\prime, m^\prime}(r,t)}{\partial r}+ \left(\beta^K_{(l,m),(l^\prime, m^\prime)}-\alpha^K_{(l,m),(l^\prime, m^\prime)}\right) \frac{u_{l^\prime, m^\prime}(r,t)}{r} \right)
\end{align}
where we have defined 
\begin{align}
    \alpha^K_{(l,m),(l^\prime, m^\prime)} &= \int Y_{l,m}^*(\Omega) \alpha_K(\Omega) Y_{l^\prime, m^\prime}(\Omega) d\Omega, \\
    \beta^K_{(l,m),(l^\prime, m^\prime)} &= \int Y_{l,m}^*(\Omega) \beta_K(\Omega)  Y_{l^\prime, m^\prime}(\Omega) d\Omega.
\end{align}
The matrix elements $\alpha^K_{(l,m),(l^\prime, m^\prime)}$ and $\beta^K_{(l,m),(l^\prime, m^\prime)}$ can be evaluated analytically by using the relations given by Eqs.~(\ref{x_sph})-(\ref{dz_sph}) or numerically by Lebedev quadrature~\cite{lebedev1975values}, which approximates the surface integral of a function over a three-dimensional sphere~\footnote{\url{https://cbeentjes.github.io/files/Ramblings/QuadratureSphere.pdf}}. Lebedev points of order $3-131$, needed to perform Lebedev quadrature, are found at~\footnote{\url{https://people.sc.fsu.edu/~jburkardt/datasets/sphere_lebedev_rule/sphere_lebedev_rule.html}}.

Thus, 
\begin{align}
i \dot{u}_{l,m}(r,t) = &-\frac{1}{2}\nabla_l u_{l,m}(r,t) + V(r)u_{l,m}(r,t) \nonumber \\ 
&-iA(t)\sum_{l^\prime,m^\prime}  \alpha^K_{(l,m),(l^\prime, m^\prime)} \frac{\partial u_{l^\prime, m^\prime}(r,t)}{\partial r} \nonumber \\
&-iA(t)\sum_{l^\prime,m^\prime} \left(\beta^K_{(l,m),(l^\prime, m^\prime)}-\alpha^K_{(l,m),(l^\prime, m^\prime)}\right) \frac{u_{l^\prime, m^\prime}(r,t)}{r} \label{radial_eoms_vel_gauge_dip_lin_pol}. 
\end{align}

We proceed as in Sec.~(\ref{sph_discrete_tise}) and introduce the scaling 
\begin{equation}
    u_{l,m}(r(x), t) = \dot{r}^{-1/2}(x) f_{l,m}(x, t).
\end{equation}
Insertion into Eq.~(\ref{radial_eoms_vel_gauge_dip_lin_pol}) yields
\begin{align}
i \dot{f}_{l,m}(x,t) &= \left(-\frac{1}{2} \frac{1}{\dot{r}(x)} \frac{d^2}{dx^2} \frac{1}{\dot{r}(x)} +\tilde{V}_{l,m}(r(x)) \right) f_{l,m}(x,t) \nonumber \\
&-iA(t)\sum_{l^\prime,m^\prime}  \alpha^K_{(l,m),(l^\prime, m^\prime)} \frac{1}{\dot{r}(x)^{1/2}} \frac{d}{dx} \frac{1}{\dot{r}(x)^{1/2}} f_{l^\prime, m^\prime}(x,t) \nonumber \\
&-iA(t)\sum_{l^\prime,m^\prime} \left(\beta^K_{(l,m),(l^\prime, m^\prime)}-\alpha^K_{(l,m),(l^\prime, m^\prime)}\right) \frac{f_{l^\prime, m^\prime}(x,t)}{r}  \label{radial_tise_linpol_x},
\end{align}
where we have used that (see Sec.~(\ref{subsec:symmetrization}))
\begin{align}
    \dot{r}(x)^{1/2} \frac{d \psi}{dr} &= \dot{r}(x)^{-1/2} \frac{d \psi}{dx} =  \frac{1}{\dot{r}(x)^{1/2}} \frac{d}{dx} \frac{1}{\dot{r}(x)^{1/2}} f(x),
\end{align} 
and 
\begin{equation}
    \tilde{V}_{l,m}(r(x)) = \tilde{V}_l(r(x)),
\end{equation}
with $\tilde{V}_l(r(x))$ given by Eq.~(\ref{func:tilde_Vl}). Note that even though $\tilde{V}_l$ is independent of $m$, it acts on every $\Tilde{f}_{l,m}$.

Finally, discretizing Eq.~(\ref{radial_tise_linpol_x}) with the pseudospectral method yields 
\begin{align}
i \dot{\tilde{f}}_{(l,m)}(x_i,t) &= \sum_{j=1}^{N-1} \left(-\frac{1}{2}  \frac{\tilde{g}^{\prime \prime}_j(x_i)}{\dot{r}(x_i) \dot{r}(x_j)}\tilde{f}_{(l,m)}(x_j,t)\right) +   \tilde{V}_{(l,m)}(r(x_i))\tilde{f}_{(l,m)}(x_i, t)\nonumber \\
&-iA(t)\sum_{(l^\prime,m^\prime)}  \alpha^K_{(l,m),(l^\prime, m^\prime)} \sum_{j=1}^{N-1} \left(\frac{\tilde{g}^{\prime }_j(x_i)}{\dot{r}(x_i)^{1/2} \dot{r}(x_j)^{1/2}}\tilde{f}_{(l^\prime, m^\prime)}(x_j,t)\right) \nonumber \\
&-iA(t)\sum_{(l^\prime,m^\prime)}\left(\beta^K_{(l,m),(l^\prime, m^\prime)}-\alpha^K_{(l,m),(l^\prime, m^\prime)}\right) \frac{\tilde{f}_{(l^\prime, m^\prime)}(x_i,t)}{r(x_i)} \label{discrete_radial_tise_linpol_x},
\end{align}
where $\tilde{g}^\prime_j(x_i)$ and $\tilde{g}_j^{\prime \prime}(x_i)$ are given by Eq.~(\ref{dg_dx_dg_tilde}) and Eq.~(\ref{ddg_ddx_ddg_tilde}), respectively, and 
\begin{equation}
    \tilde{f}_{l,m}(x_i,t) = \frac{f_{l,m}(x_i,t)}{P_N(x_i)}.
\end{equation}

Next, we define the following mapping of the angular momentum quantum numbers $(l,m)$
\begin{align}
    (l,m) &\rightarrow I(l,m) \equiv I \label{map:lm_to_I} \\
    (0,0) &\rightarrow 0 \nonumber \\
    (1,-1) &\rightarrow 1 \nonumber \\
    (1,0) &\rightarrow 2 \nonumber \\
    (1,1) &\rightarrow 3 \nonumber \\
    (2,-2) &\rightarrow 4 \nonumber \\
    &\vdots \nonumber  \\
    (l_{\text{max}}, l_{\text{max}}) &\rightarrow (l_{\text{max}}+1)^2-1 \equiv N_{l,m}  \nonumber .
\end{align}
Also, let 
\begin{equation}
    \tilde{f}_{(l,m)}(x_i, t) = \tilde{f}_{(l,m),i} = \tilde{f}_{I,i}, \ \ i=1,\cdots,N-1. 
\end{equation}
Then, we can define the matrix $\tilde{\mathbf{F}} \in \mathbb{C}^{N_{l,m} \times (N-1)}$
\begin{equation}
    \mathbf{\Tilde{F}} = 
    \begin{bmatrix}
    \tilde{f}_{0,1} & \cdots & \tilde{f}_{0,N-1} \\
    \tilde{f}_{1,1} & \cdots & \tilde{f}_{1,N-1} \\
    \vdots & \cdots & \vdots \\
    \tilde{f}_{N_{l,m},1} & \cdots & \tilde{f}_{N_{l,m},N-1}
    \end{bmatrix} ,
\end{equation}
and the vector $\tilde{\mathbf{f}} \in \mathbb{C}^{N_{l,m}(N-1)}$  obtained by stacking the rows of $\Tilde{\mathbf{F}}$ on top of one another
\begin{equation}
    \tilde{\mathbf{f}} \equiv 
    \begin{bmatrix}
    \tilde{f}_{0,1} \\
    \vdots \\
    \tilde{f}_{0,N-1} \\
    \tilde{f}_{1,1} \\
    \vdots \\
    \tilde{f}_{1,N-1} \\
    \vdots \\
    \tilde{f}_{N_{l,m},1} \\
    \vdots \\
    \tilde{f}_{N_{l,m},N-1}
    \end{bmatrix}.
\end{equation}
Then, Eq.~(\ref{discrete_radial_tise_linpol_x}), can be written on matrix-vector form as 
\begin{align}
    i \dot{\tilde{\mathbf{f}}} = H(t) \tilde{\mathbf{f}} \label{discrete_tise_matrix_vector_form},  
\end{align}
with the time-dependent Hamilton matrix given by
\begin{align}
    H(t) &= T + I_{N_{l,m}} \otimes V \nonumber \\
    &- i A(t) \left( \alpha^K \otimes D_1 + (\beta^K-\alpha^K) \otimes \text{diag}\left( \{ \frac{1}{r(x_i)} \}_{i=1}^{N-1}\right) \right),
\end{align}
where $I_{N_{l,m}} \in \mathbb{R}^{N_{l,m} \times N_{l,m}}$ is the identity matrix$, \otimes$ denotes the Kronecker product and 
\begin{equation}
    D_{1} \equiv \left[ \frac{\tilde{g}^{\prime }_j(x_i)}{\dot{r}(x_i)^{1/2} \dot{r}(x_j)^{1/2}} \right].
\end{equation}
Moreover, $T$ is the block diagonal matrix 
\begin{align}
    T &= \text{diag}\left( \{ T_{(l,m)} \} \right), \\
    T_{(l,m)} &= T_l  \label{matrix:T_lm}, 
\end{align}
where  $T_l$ is given by Eq.~(\ref{Tl_matrix}). Here we also note that even though $T_l$ is independent of $m$, it acts on every $\Tilde{f}_{l,m}$.


Applying the Crank-Nicolson method (see Sec.(\ref{sec:crank_nicolson})) to Eq.~(\ref{discrete_tise_matrix_vector_form}) yields 
\begin{equation}
    \left(I + \frac{i \Delta t}{2} H(t_{n+\frac{1}{2}}) \right) \Tilde{\mathbf{f}}^{n+1} = \left(I - \frac{i \Delta t}{2} H(t_{n+\frac{1}{2}}) \right) \Tilde{\mathbf{f}}^{n} \label{CN_sph_discrete},
\end{equation}
where $I = I_{N_{l,m}} \otimes I_{N-1} \in \mathbb{R}^{N_{l,m}(N-1) \times N_{l,m}(N-1)}$. The linear system given by Eq.~(\ref{CN_sph_discrete}) is too large to solve efficiently with direct methods (for practical grid sizes/relevant $N_{l,m}$ and $N$). 


However,  we can evaluate the matrix-vector product $H(t)\Tilde{\mathbf{f}}$ without forming $H(t)$ explicitly. Using the mapping~(\ref{map:lm_to_I}), the right-hand side of Eq.~(\ref{discrete_radial_tise_linpol_x}) can be written as
\begin{align}
\left(H(t)\Tilde{\mathbf{f}} \right)_{I,i} &= -\frac{1}{2} \left(\sum_{j=1}^{N-1} D_{2, i,j}\tilde{f}_{I, j} \right) +   \tilde{V}_I(r(x_i))\tilde{f}_{I, i}\nonumber \\
&-iA(t) \left(\sum_{J=0}^{N_{l,m}}  \alpha^K_{I,J} \left(\sum_{j=1}^{N-1} D_{1, i,j}\tilde{f}_{J, j} \right) + \sum_{J=0}^{N_{l,m}}\left(\beta^K_{I,J}-\alpha^K_{I,J}\right) \frac{\tilde{f}_{J,i}}{r(x_i)} \right) \label{discrete_radial_tise_linpol_x_mapped}.
\end{align}

Thus, we can use an iterative method, in particular, we use the BiCGSTAB method~\cite{barrett1994templates}.  

When using the BiCGSTAB method in SciPy, we have found that it is crucial to provide a preconditioner. 

In Ref.~\cite{Stefanos_TDCIS_atoms_2022} it is argued that the equations of motion in spherical coordinates become stiff due to the appearance/presence of the centrifugal term
\begin{equation}
    C_l(r) = \frac{l(l+1)}{r^2},
\end{equation}
in the Laplacian/kinetic energy operator/kinetic energy matrix $T$.

We use the preconditioner 
\begin{equation}
    M = \text{blockdiag}(M^{(0,0)}, \cdots, M^{(l_{\text{max}},l_{\text{max}})}),
\end{equation}
where 
\begin{equation}
    M^{(l,m)} = \left(I + \frac{i \Delta t}{2} T_{(l,m)} \right)^{-1},
\end{equation}
where $T_{(l,m)}$ is the matrix defined by Eq.~(\ref{matrix:T_lm}). One can also include the potential $\tilde{V}_{l,m} = \Tilde{V}_{l}$ in the preconditioner matrix, i.e., 
\begin{equation}
    \Tilde{M}^{(l,m)} = \left(I + \frac{i \Delta t}{2} \left( T_{(l,m)} + \tilde{V}_{l,m} \right) \right)^{-1}.
\end{equation} 
We have not found that there is any significant benefit to using $\Tilde{M}^{(l,m)}$ over $M^{(l,m)}$. 


The matrices $\{ M^{(l,m)} \}$ (or $\{ \tilde{M}^{(l,m)} \}$) can be computed before the propagation by direct inversion and stored in memory, as long as the number of grid points, $N$, is within reason. Typically we have $N \in [100, 1500]$.

\appendix
\section{Spherical coordinates}
We use spherical coordinates $\mathbf{r} = (r, \theta, \phi)$, related to Cartesian coordinates by
\label{sec:sph_coords}
\begin{align}
x &= r \sin \theta \cos \phi, \\
y &= r \sin \theta \sin \phi, \\
z &= r \cos \theta,
\end{align}
where $r \in [0,\infty)$, $\theta \in [0,\pi]$ and $\phi \in [0,2\pi)$, and the volume element is given by  $dV = r^2 \sin \theta  dr d\theta d\phi$.

The Cartesian derivative operators are given by 
\begin{align}
\frac{\partial}{\partial x} &= \cos{\phi} \sin{\theta}\frac{\partial}{\partial r} 
    + \frac{1}{r} \left( - \frac{\sin{\phi}}{\sin{\theta}}\frac{\partial}{\partial \phi} 
    + \cos{\phi}\cos{\theta}\frac{\partial}{\partial \theta} \right) \label{sph_dx}\\
    \frac{\partial}{\partial y} &= \sin{\phi} \sin{\theta}\frac{\partial}{\partial r} 
    + \frac{1}{r} \left( \frac{\cos{\phi}}{\sin{\theta}}\frac{\partial}{\partial \phi} 
    + \sin{\phi}\cos{\theta}\frac{\partial}{\partial \theta} \right) \label{sph_dy} \\
    \frac{\partial}{\partial z} &= \cos{\theta}\frac{\partial}{\partial r} 
    - \frac{\sin{\theta}}{r}\frac{\partial}{\partial \theta} \label{sph_dz},    
\end{align}
and the Laplacian is given by 
\begin{align}
\nabla^2 &= \frac{1}{r^2} \frac{\partial}{\partial r}\left( r^2 \frac{\partial}{\partial r} \right) + \frac{1}{r^2} \left[\frac{1}{\sin(\theta)}\frac{\partial}{\partial \theta}\left(\sin(\theta) \frac{\partial}{\partial \theta}\right) +\frac{1}{\sin^2(\theta)}\frac{\partial^2}{\partial \phi^2}\right] \\
&= \frac{1}{r} \frac{\partial^2}{\partial r^2} r - \frac{\hat{L}^2}{r^2}.
\end{align}

\section{Spherical harmonics}
\label{sec:sph_harmonics}
We use the convention where the spherical harmonics $Y_l^m(\theta, \phi)$ are defined as
\begin{equation}
Y_l^m(\theta, \phi) = (-1)^m\sqrt{\frac{2l+1}{4\pi}\frac{(l-m)!}{(l+m)!}}P_l^m(\cos\theta)e^{im\phi},
\end{equation}
where $P_l^m(x)$ are the associated Legendre functions
\begin{equation}
P_l^m(x) = (1-x^2)^{|m|/2} \left(\frac{d}{dx}\right)^{|m|} P_l(x),
\end{equation}
and $P_l(x)$ are the Legendre polynomials
\begin{equation}
P_l(x) = \frac{1}{2^l l!} \left( \frac{d}{dx} \right)^l (x^2-1)^l.
\end{equation}
In the following we use the notation 
\begin{equation}
    Y_l^m(\theta, \phi) = Y_l^m(\Omega)
\end{equation}
where $\Omega = (\theta, \phi)$.

\subsection{Properties}
\begin{equation}
    \int (Y_l^m)^*(\Omega)Y_{l^\prime}^{m^\prime}(\Omega) d\Omega = \delta_{l l^\prime}\delta_{m m^\prime}
\end{equation}

\begin{equation}
    (Y_l^m)^*(\Omega) = (-1)^{m}Y_l^{-m}(\Omega)
\end{equation}

\begin{equation}
    \sum_{m=-l}^l |Y_l^m(\Omega)|^2 = \frac{2l+1}{4 \pi}
\end{equation}

The product of two spherical harmonics can be expressed as 
\begin{equation}
    Y_{l^\prime,m^\prime}(\Omega)Y_{l,m}(\Omega) = \sqrt{\frac{(2l^\prime+1)(2l+1)}{4\pi}}\sum_{L=0}^{\infty}\sum_{M=-L}^{L}(-1)^M\sqrt{2L+1}
    \begin{pmatrix}
        l^\prime & l & L \\
        m^\prime & m & -M 
    \end{pmatrix}
    \begin{pmatrix}
        l^\prime & l & L \\
        0 & 0 & 0 
    \end{pmatrix}
    Y_{L,M}(\Omega),
\end{equation}
where $\begin{pmatrix} l^\prime & l & L \\ m^\prime & m & -M \end{pmatrix}$ are the Wigner-$3j$ symbols~\cite{arfken2011mathematical}.


Trigonometric functions can be written in terms of spherical harmonics, and their actions on spherical harmonics can then be evaluated using 
the product rule for two spherical harmonics. Some important relations are
\begin{align}
    \cos{\theta} &= 2\sqrt{\frac{\pi}{3}}Y_{1,0}(\Omega) \\
    \sin{\theta}e^{\pm i\phi} &= 2\sqrt{\frac{2\pi}{3}}Y_{1,\pm 1}(\Omega) \\
    \sin{\theta}\cos{\phi} &= \sqrt{\frac{2\pi}{3}}\Bigl[ Y_{1,-1}(\Omega) - Y_{1,1}(\Omega) \Bigr] \\
    \sin{\theta}\sin{\phi} &= \sqrt{\frac{2\pi}{3}}i\Bigl[ Y_{1,-1}(\Omega) + Y_{1,1}(\Omega) \Bigr]
\end{align}

\subsection{The effect of some selected operators on the spherical harmonics}

\begin{align}
    \frac{\partial}{\partial \phi}Y_{l,m}(\Omega) &= imY_{l,m}(\Omega), \\
    \frac{\partial}{\partial \theta}Y_{l,m}(\Omega) &= m \frac{\cos{\theta}}{\sin{\theta}}Y_{l,m}(\Omega) + \sqrt{(l-m)(l+m+1)}e^{-i\phi}Y_{l,m+1}(\Omega)
\end{align}

The action of the Cartesian position and derivative operators on the spherical harmonics are given by
\begin{align}
xY_{l,m} &= \frac{1}{2}r\Bigl[\Bigr.b_{l-1,-m-1}Y_{l-1,m+1} - b_{l,m}Y_{l+1,m+1} - b_{l-1,m-1}Y_{l-1,m-1} + b_{l,-m}Y_{l+1,m-1}\Bigl. \Bigr], \label{x_sph}\\
    yY_{l,m} &= \frac{i}{2}r\Bigl[\Bigr.b_{l-1,-m-1}Y_{l-1,m+1} - b_{l,m}Y_{l+1,m+1} + b_{l-1,m-1}Y_{l-1,m-1} - b_{l,-m}Y_{l+1,m-1}\Bigl. \Bigr], \label{y_sph} \\
    zY_{l,m} &= r\Bigl[a_{l,m}Y_{l+1,m} + a_{l-1,m}Y_{l-1,m}\Bigr] \label{z_sph},
\end{align}
and
\begin{align}
\frac{\partial}{\partial x} Y_{l,m} &= \cos{\phi}\sin{\theta}Y_{l,m}\frac{\partial}{\partial r} \nonumber \\
&+ \frac{1}{2} \left[\cos\theta \left( c_{l,m}Y_{l,m+1} - c_{l,m-1}Y_{l,m-1} \right) + m \sin \theta (e^{-i \phi}-e^{i\phi}) Y_{l,m} \right] \frac{1}{r}, \label{dx_sph} \\
\frac{\partial}{\partial y} Y_{l,m} &= \sin{\phi}\sin{\theta}Y_{l,m}\frac{\partial}{\partial r} \nonumber \\
&+ \frac{i}{2} \left[-\cos\theta \left( c_{l,m}Y_{l,m+1} + c_{l,m-1}Y_{l,m-1} \right) + m \sin \theta (e^{-i \phi}+e^{i\phi}) Y_{l,m} \right] \frac{1}{r}, \label{dy_sph}\\
\frac{\partial}{\partial z} Y_{l,m} &= \bigl(a_{l,m}Y_{l+1,m} + a_{l-1,m}Y_{l-1,m}\bigr)\frac{\partial}{\partial r} \nonumber \\
&+ \Bigl[ -la_{l,m}Y_{l+1,m}  + (l+1)a_{l-1,m}Y_{l-1,m}\Bigr] \frac{1}{r}, \label{dz_sph}  
\end{align}
where we have defined 
\begin{align}
a_{l,m} &= \sqrt{\frac{(l+1)^2-m^2}{(2l+1)(2l+3)}}, \\
b_{l,m} &= \sqrt{\frac{(l+m+1)(l+m+2)}{(2l+1)(2l+3)}}, \\
c_{l,m} &= \sqrt{l(l+1) - m(m+1)}.
\end{align}

\section{Miscellaneous derivations}
\label{sec:misc_derivations}
\subsection{Chain rule}
\label{subsec:chain_rule}
Assume that $\psi(r) = \psi(r(x))$. By the chain rule we find that
\begin{equation}
    \frac{d \psi}{dx} = \frac{d \psi}{dr} \frac{dr}{dx} = \frac{d \psi}{dr} \dot{r}(x),
\end{equation}
such that
\begin{equation}
    \frac{d \psi}{dr} = \frac{1}{\dot{r}(x)} \frac{d \psi}{dx} \label{dpsi_dr_of_x}.
\end{equation}
A second application of the chain rule gives that
\begin{align}
    \frac{d^2 \psi}{dx^2} = \frac{d^2\psi}{dr^2} \dot{r}(x)^2 + \frac{d \psi}{dr} \ddot{r}(x),
\end{align}
and using Eq.(\ref{dpsi_dr_of_x}) we obtain
\begin{equation}
    \frac{d^2 \psi}{dr^2} = \frac{1}{\dot{r}(x)^2} \frac{d^2 \psi}{dx^2} - \frac{\ddot{r}(x)}{\dot{r}(x)^3} \frac{d \psi}{dx}. \label{ddpsi_ddrx}
\end{equation}

\subsection{Symmetrization}
\label{subsec:symmetrization}

Introducing the scaling $\psi(x) = \dot{r}(x)^{-1/2} f(x)$ we find that
\begin{align}
    \frac{d \psi}{dr} &= \frac{1}{\dot{r}(x)}\frac{d \psi}{dx} \\
    &= \frac{1}{\dot{r}(x)} \left(\dot{r}(x)^{-1/2} \frac{d f}{dx} - \frac{1}{2} \dot{r}(x)^{-3/2}\ddot{r}(x) f(x) \right) \nonumber \\
    &= \dot{r}(x)^{-3/2} \frac{df}{dx} - \frac{1}{2}\dot{r}(x)^{-5/2}\ddot{r}(x)f(x) \nonumber \\
    &= \dot{r}(x)^{-1/2} \left( \frac{1}{\dot{r}(x)} \frac{df}{dx} - \frac{\ddot{r}(x)}{2\dot{r}(x)^2} f(x) \right) \nonumber \\
    &= \dot{r}(x)^{-1/2} \left( \frac{1}{\dot{r}(x)^{1/2}} \frac{d}{dx} \frac{1}{\dot{r}(x)^{1/2}} f(x) \right).
\end{align}
Furthermore, computing $\frac{d}{dx} \left( \dot{r}(x)^{-1/2}f(x) \right)$ and $\frac{d^2}{dx^2} \left( \dot{r}(x)^{-1/2}f(x) \right)$ we find that  
\begin{align}
    \frac{1}{\dot{r}(x)^2} \frac{d^2 \psi}{dx^2} - \frac{\ddot{r}(x)}{\dot{r}(x)^3} \frac{d \psi}{dx} &= \frac{1}{\dot{r}(x)^{5/2}}\frac{d^2 f}{dx^2} -  \frac{2\ddot{r}(x)}{\dot{r}(x)^{7/2}} \frac{df}{dx} \nonumber \\ 
    &- \frac{\dddot{r}(x)}{2 \dot{r}(x)^{7/2}} f(x)  + \frac{5\ddot{r}(x)^2}{4 \dot{r}(x)^{9/2}} f(x) 
\end{align}
Inserting this expression into Eq.~(\ref{SE_of_x}) (and multiplying on both sides with $\dot{r}^{1/2}$) we have that
\begin{equation}
    -\frac{1}{2} \left( \frac{1}{\dot{r}(x)^{2}}\frac{d^2 f}{dx^2} -  \frac{2\ddot{r}(x)}{\dot{r}(x)^{3}} \frac{df}{dx} - \frac{\dddot{r}(x)}{2 \dot{r}(x)^{3}} f(x)  + \frac{5\ddot{r}(x)^2}{4 \dot{r}(x)^{4}} f(x) \right) + V_l(r(x)) f(x) = \epsilon f(x).
\end{equation}
Now, 
\begin{equation}
    \frac{1}{\dot{r}(x)} \frac{d^2}{dx^2} \frac{1}{\dot{r}(x)} f(x) = \frac{1}{\dot{r}(x)^{2}}\frac{d^2 f}{dx^2} -  \frac{2\ddot{r}(x)}{\dot{r}(x)^{3}} \frac{df}{dx} \nonumber - \frac{\dddot{r}(x)}{ \dot{r}(x)^{3}} f(x)  + \frac{2\ddot{r}(x)^2}{ \dot{r}(x)^{4}} f(x)
\end{equation}
Then, we observe that 
\begin{align}
    &\frac{1}{\dot{r}(x)^{2}}\frac{d^2 f}{dx^2} -  \frac{2\ddot{r}(x)}{\dot{r}(x)^{3}} \frac{df}{dx} - \frac{\dddot{r}(x)}{2 \dot{r}(x)^{3}} f(x)  + \frac{5\ddot{r}(x)^2}{4 \dot{r}(x)^{4}} f(x)  \nonumber \\
    = &\frac{1}{\dot{r}(x)} \frac{d^2}{dx^2} \frac{1}{\dot{r}(x)} f(x) + \frac{\dddot{r}(x)}{2\dot{r}(x)^3}f(x) - \frac{3\ddot{r}(x)^2}{4\dot{r}(x)^4}f(x) \\
    = &\frac{1}{\dot{r}(x)} \frac{d^2}{dx^2} \frac{1}{\dot{r}(x)} f(x) + \left(\frac{2\dddot{r}(x)\dot{r}(x)-3\ddot{r}(x)^2}{4\dot{r}(x)^4} \right)f(x),
\end{align}
where we note that the term 
\begin{equation}
    h(x) \equiv \frac{2\dddot{r}(x)\dot{r}(x)-3\ddot{r}(x)^2}{4\dot{r}(x)^4}, 
\end{equation}
acts like an additional potential. 

For the particular mapping given by Eq.~(\ref{x_to_r}), one finds that
\begin{equation}
    2\dddot{r}(x)\dot{r}(x)-3\ddot{r}(x)^2 = 0.
\end{equation}
In that case, 
\begin{equation}
    \frac{1}{\dot{r}(x)^{2}}\frac{d^2 f}{dx^2} -  \frac{2\ddot{r}(x)}{\dot{r}(x)^{3}} \frac{df}{dx} - \frac{\dddot{r}(x)}{2 \dot{r}(x)^{3}} f(x)  + \frac{5\ddot{r}(x)^2}{4 \dot{r}(x)^{4}} f(x)  = \frac{1}{\dot{r}(x)} \frac{d^2}{dx^2} \frac{1}{\dot{r}(x)} f(x).
\end{equation}


\printbibliography

\end{document}